\documentclass[orivec]{llncs}

\usepackage{url}
\usepackage{wrapfig}
\usepackage{amsmath,amssymb}
\usepackage{listings}

\usepackage{stmaryrd} 

\usepackage[bookmarks,linkcolor=red,citecolor=blue,urlcolor=blue,colorlinks,breaklinks,bookmarksnumbered,bookmarksopen]{hyperref}

\usepackage{ded}
\usepackage{basics}
\usepackage{local}
\newcommand{\defemph}[1]{{\bf #1}}

\begin{document}

\title{A Query Language for Formal Mathematical Libraries}
\author{Florian Rabe}
\institute{Jacobs University Bremen, Germany}
\maketitle

\begin{abstract}
One of the most promising applications of mathematical knowledge management is search:
Even if we restrict attention to the tiny fragment of mathematics that has been formalized, the amount exceeds the comprehension of an individual human.

Based on the generic representation language MMT, we introduce the mathematical query language QMT: It combines simplicity, expressivity, and scalability while avoiding a commitment to a particular logical formalism.
QMT can integrate various search paradigms such as unification, semantic web, or XQuery style queries, and QMT queries can span different mathematical libraries.

We have implemented QMT as a part of the MMT API. This combination provides a scalable indexing and query engine that can be readily applied to any library of  mathematical knowledge.
While our focus here is on libraries that are available in a content markup language, QMT naturally extends to presentation and narration markup languages.
\end{abstract}

\section{Introduction and Related Work}\label{sec:intro}
  Mathematical knowledge management applications are particularly strong at large scales, where automation can be significantly superior to human intuition.
This makes search and retrieval pivotal MKM applications: The more the amount of mathematical knowledge grows, the harder it becomes for users to find relevant information.
Indeed, even expert users of individual libraries can have difficulties reusing an existing development because they are not aware of it.
Therefore, this question has received much attention.

\defemph{Object query languages} augment standard text search with phrase queries that match mathematical operators and with wild cards that match arbitrary mathematical expressions.
Abstractly, an object query engine is based on an index, which is a set of pairs $(l,o)$ where $o$ is an object and $l$ is the location of $o$.
The index is built from a collection of mathematical documents, and the result of an object query is a subset of the index.
The object model is usually based on presentation MathML and/or content MathML/OpenMath \cite{mathml,openmath}, but importers can be used to index other formats such as LaTeX. 
Examples for object query languages and engines are given in \cite{dlmf,mathfind,egomath,mathwebsearch,mias}.
A partial overview can be found in \cite{mias}.
A central question is the use of wild cards. An example language with complex wild cards is given in \cite{altamimi_youssef_querylanguage}.
Most generally, \cite{mathwebsearch} uses unification queries that return all objects that can be unified with the query.

\defemph{Property query languages} are similar to object query languages except that both the index and the query use relational information that abstracts from the mathematical objects. For example, the relational index might store the toplevel symbol of every object or the ``used-in'' relation between statements.
This approximates an object index, and many property queries are special cases of object queries.
But property queries are simpler and more efficient, and they still cover many important examples.
Such languages are given in \cite{mowgli_retrieval,mowgli_retrieval2} and \cite{mizar_retrieval} based on the Coq and Mizar libraries, respectively.

\defemph{Compositional query languages} focus on a complex language of query expressions that are evaluated compositionally. The atomic queries are provided by the elements of the queried library.
SQL \cite{sql} uses $n$-ary relations between elements, and query expressions use the algebra of relations.
The SPARQL \cite{sparql} data model is RDF, and queries focus on unary and binary predicates on a set of URIs of statements. This could serve as the basis for mathematics on the semantic web.
Both data models match bibliographical meta-data and property-based indices and could also be applied to the results of object queries (seen as sets of pairs); but they are not well-suited for expressions.
The XQuery \cite{xquery} data model is XML, and query expressions are centered around operations on lists of XML nodes. This is well-suited for XML-based markup languages for mathematical documents and expressions and was applied to {\sc OMDoc} \cite{omdoc} in \cite{tntbase}. In \cite{KRZ:mmttnt:10}, the latter was combined with property queries.
Very recently \cite{queryingproofs} gave a compositional query language for hiproof proof trees that integrates concepts from both object and property queries.

A number of \defemph{individual libraries} of mathematics provide custom query functionality.
Object query languages are used, for example, in \cite{activemath_retrieval} for Activemath or in Wolfram$|$Alpha.
Most interactive proof assistants permit some object or property queries, primarily to search for theorems that are applicable to a certain goal, e.g., Isabelle, Coq, and Matita.
\cite{mizar_retrieval2} is notable for using automated reasoning to prepare an index of all Mizar theorems.
\medskip

It is often desirable to combine several of the above formalisms in the same query.
Therefore, we have designed the query language QMT with the goal of permitting as many different query paradigms as possible.
QMT uses a simple kernel syntax in which many advanced query paradigms can be defined.
This permits giving a formal syntax, a formal semantics, and a scalable implementation, all of which are presented in this paper.

QMT is grounded in the {\mmt} language (Module System for Mathematical Theories) \cite{RK:mmt:10}, a scalable, modular representation language for mathematical knowledge.
It is designed as a scalable trade-off between (i) a logical framework with formal syntax and semantics and (ii) an MKM framework that does not commit to any particular formal system.
Thus, {\mmt} permits both adequate representations of virtually any formal system as well as the implementation of generic MKM services.
We implement QMT on top of our {\mmt} system, which provides a flexible and scalable query API and query server.

Our design has two pivotal strengths. Firstly, QMT can be applied to the libraries of any formal system that is represented as {\mmt}. Queries can even span libraries of different systems.
Secondly, QMT queries can make use of other {\mmt} services. For example, queries can access the inferred type and the presentation of a found expression, which are computed dynamically.
\medskip

We split the definition of QMT into two parts.
Firstly, Sect.~\ref{sec:query} defines QMT signatures in general and then the syntax and semantics of QMT for an arbitrary signature.
Secondly, Sect.~\ref{sec:mmtquery} describes a specific QMT signature that we use for {\mmt} libraries.
Our implementation, which is based on that signature, is presented in Sect.~\ref{sec:impl}.

\section{The QMT Query Language}\label{sec:query}
  \subsection{Syntax}

\begin{figure}[t]
\begin{center}
\begin{tabular}{|l|l|}\hline
Declaration & Intended Semantics \\
\hline
base type $a$          & a set of objects \\
concept symbol $c$     & a subset of a base type \\
relation symbol $r$    & a relation between two base types \\
function symbol $f$    & a sorted first-order function \\
predicate symbol $p$   & a sorted first-order predicate \\
\hline
\end{tabular}
\medskip

\begin{tabular}{|l|l|}\hline
Kind of Expression & Intended Semantics \\
\hline
Type $T$ & a set \\
Query $Q:T$ & an element of $T$ \\
 \tb element query $Q:t$   & \tb an element of $t$ \\
 \tb set query $Q:\set{t}$ & \tb a subset of $t$ \\
Relation $R<\fp{a,a'}$ & a relation between $a$ and $a'$ \\
Proposition $F$ & a boolean truth value\\
\hline
\end{tabular}
\caption{QMT Notions and their Intuitions}\label{fig:intuitions}
\end{center}
\vspace{-2em}
\end{figure}

Our syntax arises by combining features of sorted first-order logic -- which leads to very intuitive expressions -- and of description logics -- which leads to efficient evaluations. Therefore, our \defemph{signatures} $\Sigma$ contain five kinds of declarations as given in Fig.~\ref{fig:intuitions}.

For a given signature, we define four kinds of \defemph{expressions}: types $T$, relations $R$, propositions $F$, and typed queries $Q$ as listed in Fig.~\ref{fig:intuitions}. The grammar for signatures and expressions is given in Fig.~\ref{fig:grammar}.
\medskip

The intuitions for most expression formation operators can be guessed easily from their notations. In the following we will discuss each in more detail.

Regarding \defemph{types} $T$, we use product types and power type. However, we go out of our way to avoid arbitrary nestings of type constructors. Every type is either a product $t=\qp{a_1,\ldots,a_n}$ of base types $a_i$ or the power type $\set{t}$ of such a type. Thus, we are able to use the two most important type formation operators in the context of querying: product types arise when a query contains multiple query variables, and power types arise when a query returns multiple results. But at the same time, the type system remains very simple and can be treated as essentially first-order.

\begin{figure}[t]
\begin{center}
\begin{tabular}{|llcl|}\hline
Signatures     & $\Sigma$  & $\bnfas$ & $\cdot 
  \bnfalt \Sigma,\;a:\btp
  \bnfalt \Sigma,\;c < a
  \bnfalt \Sigma,\;r < \fp{a, a}$ \\
  &&& $\bnfalt \Sigma,\;f:\fp{T,\ldots, T}\arr T
  \bnfalt \Sigma,\;p:\fp{T,\ldots,T}\arr\prop$ \\
Contexts       & $\Gamma$  & $\bnfas$ & $\cdot \bnfalt \Gamma,x:T$ \\
\hline
Simple Types   & $t$       & $\bnfas$ & $\qp{a,\ldots, a}$ \\
General Types  & $T$       & $\bnfas$ & $t \bnfalt \set{t}$ \\
Relations      & $R$       & $\bnfas$ & $r \bnfalt R^{-1} \bnfalt \tc{R} \bnfalt R;R \bnfalt R\cup R\bnfalt R\cap R\bnfalt R\sm R$\\
Propositions   & $F$       & $\bnfas$ & $p(Q,\ldots,Q) \bnfalt \neg F \bnfalt F\wedge F\bnfalt \forall x\in Q.F(x)$ \\
Queries        & $Q$       & $\bnfas$ & $c \bnfalt x \bnfalt f(Q,\ldots,Q) \bnfalt\qpi{Q,\ldots,Q}\bnfalt Q_i$ \\
                         &&& $\bnfalt \rel{R}{Q} \bnfalt \bigcup_{x\in Q}Q(x) \bnfalt \{x\in Q|F(x)\}$ \\
\hline
\end{tabular}
\caption{The Grammar for Query Expressions}\label{fig:grammar}
\end{center}
\vspace{-2em}
\end{figure}

Regarding \defemph{relations}, we provide the common operations from the calculus of binary relations: dual/inverse $R^{-1}$, transitive closure $\tc{R}$, composition $R;R'$, union $R\cup R'$, intersection $R\cap R'$, and difference $R\sm R'$. Notably absent is the absolute complement operation $R^\mathtt{C}$; it is omitted because its semantics can not be computed efficiently in general. Note that the operation $R^{-1}$ is only necessary for atomic $R$: For all other cases, we can put $(\tc{R})^{-1}=\tc{(R^{-1})}$, $(R;R')^{-1}={R'}^{-1};R^{-1}$, and $(R\ast R')^{-1}=R^{-1}\ast {R'}^{-1}$ for $\ast\in\{\cup,\cap,\sm\}$.

Regarding \defemph{propositions}, we use the usual constructors of classical first-order logic: predicates, negation, conjunction, and universal quantification. As usual, the other constructors are definable. However, there is one specialty: The quantification $\forall x\in Q.F(x)$ does not quantify over a type $t$; instead, it is relativized by a query result $Q:\set{t}$. This specialty is meant to support efficient evaluation: The extension of a base type is usually much larger than that of a query, and it may not be efficiently computable or not even finite.

Regarding \defemph{queries}, our language combines intuitions from description and first-order logic with an intuitive mathematical notation. Constants $c$, variables $x$, and function application are as usual for sorted first-order logic. $\qpi{Q^1,\ldots,Q^n}$ for $n\in\N$ and $Q_i$ for $i=1,\ldots,n$ denote tupling and projection. $\rel{R}{Q}$ represents the image of the object given by $Q$ under the relation given by $R$. $\bigcup_{x\in Q}Q'(x)$ denotes the union of the family of queries $Q'(x)$ where $x$ runs over all objects in the result of $Q$. Finally, $\{x\in Q|F(x)\}$ denotes comprehension on queries, i.e., the objects in $Q$ that satisfy $F$. Just like for the universal quantification, all bound variables are relativized to a query result to support efficient evaluation.

\begin{fignd}{sig}{Well-Formed Signatures}
\ianc{n \text{ not declared in } \Sigma}{n\nin\Sigma}{} \tb\tb
\ianc{}{\issig{\cdot}}{} \tb\tb
\ibnc{\issig{\Sigma}}{a\nin\Sigma}{\issig{\Sigma,\,a:\btp}}{} \\
\icnc{\issig{\Sigma}}{c\nin\Sigma}{a:\btp\minn\Sigma}{\issig{\Sigma,\,c<a}}{}\tb\tb
\icnc{\issig{\Sigma}}{r\nin\Sigma}{\big(a_i:\btp\minn\Sigma\big)_{i=1}^2}{\issig{\Sigma,\,r<\fp{a_1,a_2}}}{}\\
\icnc{\issig{\Sigma}}{f\nin\Sigma}{\big(\istypeS{T_i}\big)_{i=1}^{n+1}}{\issig{\Sigma,\,f:\fp{T_1,\ldots,T_n}}\arr T_{n+1}}{}\tb\tb
\icnc{\issig{\Sigma}}{p\nin\Sigma}{\big(\istypeS{T_i}\big)_{i=1}^n}{\issig{\Sigma,\,p:\fp{T_1,\ldots,T_n}\arr \prop}}{}
\end{fignd}

\begin{fignd}{typing}{Well-Formed Expressions}
\ianc{\big(a_i:\btp\minn\Sigma\big)_{i=1}^n}{\istypeS{\qp{a_1,\ldots,a_n}}}{} \tb\tb
\ianc{\big(a_i:\btp\minn\Sigma\big)_{i=1}^n}{\istypeS{\set{\qp{a_1,\ldots,a_n}}}}{}\\
\hline
\ianc{c<t\minn\Sigma}{\ofSG{c}{\set{t}}}{} \tb\tb
\ibnc{f:\fp{T_1,\ldots. T_n}\arr T\minn\Sigma}{\ofSG{Q_i}{T_i}}{\ofSG{f(Q_1,\ldots,Q_n)}{T}}{} \tb\tb
\ianc{x:T\minn\Gamma}{\ofSG{x}{T}}{} \\
\ianc{\ofSG{Q_i}{t_i}\mfor i=1,\ldots,n}{\ofSG{\qpi{Q_1,\ldots,Q_n}}{\qp{t_1,\ldots,t_n}}}{} \tb\tb
\ibnc{\ofSG{Q}{\qp{t_1,\ldots,t_n}}}{i\in\{1,\ldots,n\}}{\ofSG{Q_i}{t_i}}{} \\
\ibnc{\ofSG{Q}{\set{t}}}{\oftype{\Sigma}{\Gamma,x:t}{Q'(x)}{\set{t'}}}{\ofSG{\bigcup_{x\in Q}Q'(x)}{\set{t'}}}{} \\
\ibnc{\ofSG{Q}{t}}{\isrel{\Sigma}{R}{t}{t'}}{\ofSG{\rel{R}{Q}}{\set{t'}}}{} \tb\tb
\ibnc{\ofSG{Q}{\set{t}}}{\isprop{\Sigma}{\Gamma,x:t}{F(x)}}{\ofSG{\{x\in Q|F(x)\}}{\set{t}}}{} \\
\hline
\ianc{r<\fp{a, a'}\minn\Sigma}{\isrel{\Sigma}{r}{a}{a'}}{} \tb\tb
\ianc{\isrelS{R}{a}{a'}}{\isrelS{R^{-1}}{a'}{a}}{} \tb\tb
\ianc{\isrelS{R}{a}{a}}{\isrelS{\tc{R}}{a}{a}}{} \\
\ibnc{\isrelS{R}{a}{a'}}{\isrelS{R'}{a'}{a''}}{\isrelS{R;R'}{a}{a''}}{} \tb\tb
\icnc{\isrelS{R}{a}{a'}}{\isrelS{R'}{a}{a'}}{\ast\in\{\cup,\cap,\sm\}}{\isrelS{R\ast R'}{a}{a'}}{} \\
\hline
\ibnc{p:\fp{T_1,\ldots, T_n}\arr\prop\minn\Sigma}{\ofSG{Q_i}{T_i}}{\isprop{\Sigma}{\Gamma}{p(Q_1,\ldots,Q_n)}}{} \\
\ianc{\isprop{\Sigma}{\Gamma}{F}}{\isprop{\Sigma}{\Gamma}{\neg F}}{} \tb\tb
\ibnc{\isprop{\Sigma}{\Gamma}{F}}{\isprop{\Sigma}{\Gamma}{F'}}{\isprop{\Sigma}{\Gamma}{F\wedge F'}}{} \\
\ibnc{\ofSG{Q}{\set{t}}}{\isprop{\Sigma}{\Gamma,x:t}{F(x)}}{\isprop{\Sigma}{\Gamma}{\forall x\in Q.F(x)}}{} 
\end{fignd}

\begin{remark}
While we do not present a systematic analysis of the efficiency of QMT, we point out that we designed the syntax of QMT with the goal of supporting efficient evaluation.
In particular, this motivated our distinction between the ontology part, i.e., concept and relation symbols, and the first-order part, i.e., the function and predicate symbols.

Indeed, every concept $c<t$ can be regarded as a function symbol $c:\set{t}$, and every relation $r<\fp{a,a'}$ as a predicate symbol $r:\fp{a,a'}\arr\prop$.
Thus, the ontology symbols may appear redundant --- their purpose is to permit efficient evaluations.
This is most apparent for relations. For a predicate symbol $p:\fp{a,a'}\arr\prop$, evaluation requires a method that maps from $\sem{a}\times\sem{a'}$ to booleans. But for a relation symbol $r<\fp{a,a'}$, evaluation requires a method that returns for any $u$ all $v$ such that $(u,v)\in\sem{r}$ or all $v$ such that $(v,u)\in\sem{r}$. A corresponding property applies to concepts.

Therefore, efficient implementations of QMT should maintain indices for them that are computed a priori: hash sets for the concept symbols and hash tables for the relation symbols.
(Note that using hash tables for all relation symbols permits fast evaluation of all relation expressions $R$, which is crucial for the evaluation of queries $\rel{R}{Q}$.)
The implementation of function and predicate symbols, on the other hand, only requires plain functions that are called when evaluating a query.

Thus, it is a design decision whether a certain feature is realized by an ontology or by a first-order symbol.
By separating the ontology and the first-order part, we permit simple indices for the former and retain flexible extensibility for the latter (see also Rem.~\ref{rem:finite}).
\end{remark}

\begin{wrapfigure}{r}{8.5cm}
\vspace{-3em}
\begin{center}
\begin{tabular}{|l|l|}
\hline
Judgment & Intuition \\
\hline
$\issig{\Sigma}$ & well-formed signature $\Sigma$\\
$\istypeS{T}$ & well-formed type $T$\\
$\ofSG{Q}{T}$ & well-typed query $Q$ of type $T$\\
$\ofSG{Q}{T}$ & well-typed query $Q$ of type $T$\\
$\isrel{\Sigma}{R}{a}{a'}$ & well-typed relation $R$ between $a$ and $a'$ \\
$\isprop{\Sigma}{\Gamma}{F}$ & well-formed proposition $F$ \\
\hline
\end{tabular}
\caption{Judgments}\label{fig:judgments}
\end{center}
\vspace{-3em}
\end{wrapfigure}

Based on these intuitions, it is straightforward to define the \defemph{well-formed expressions}, i.e., the expressions that will have a denotational semantics. More formally, we use the \defemph{judgments} given in Fig.~\ref{fig:judgments} to define the well-formed expressions over a signature $\Sigma$ and a context $\Gamma$. The \defemph{rules} for these judgments are given in Fig.~\ref{fig:sig} and~\ref{fig:typing}.

In order to give some meaningful examples, we will already make use of the symbols from the MMT signature, which we will introduce in Sect. \ref{sec:mmtquery}.

\begin{example}\label{ex:include}
Consider a base type $\mmturi:\btp$ of MMT identifiers in some fixed MMT library. Moreover, consider a concept symbol $\conc{theory}<\mmturi$ giving the identifiers of all theories, and a relation symbol $\conc{includes}<\fp{\mmturi,\mmturi}$ that gives the relation ``theory $A$ directly includes theory $B$''.

Then the query $\conc{theory}$ of type $\set{\mmturi}$ yields the set of all theories. Given a theory $u$, the query $\rel{{\tc{\conc{includes}}}^{-1}}{u}$ of type $\set{\mmturi}$ yields the set of all theories that transitively include $u$.
\end{example}

\begin{example}[Continued]\label{ex:constants}
Additionally, consider a concept $\conc{constant}<\mmturi$ of identifiers of MMT constants, relation symbol $\conc{declares}<\fp{\mmturi,\mmturi}$ that relates every theory to the constants declared in it, a base type $\mmtobj:\btp$ of {\openmath} objects, a function symbol $\typeOF:\mmturi\arr\opt{\mmtobj}$ that maps each MMT constant to its type, and a predicate symbol $\occurs:\fp{\mmturi,\mmtobj}\arr\prop$ that determines whether an identifier occurs in an object.

Then the following query of type $\set{\mmturi}$ retrieves all constants that are included into the theory $u$ and whose type uses the identifier $v$:
\[\{x\in \rel{(\tc{\conc{includes}};\conc{declares})}{u} \,|\, \occurs(v,\typeOF(x))\}\]
\end{example}

\subsection{Semantics}

\begin{wrapfigure}{r}{5.5cm}
\vspace{-5em}
\begin{center}
\begin{tabular}{|l|l|}
\hline
Judgment & Semantics \\
\hline
$\istypeS{T}$                & $\sem{T}\in\Set$\\
\tb $\ofSG{Q}{t}$            & $\sema{Q}\in\sem{t}$\\
\tb $\ofSG{Q}{\set{t}}$      & $\sema{Q}\sq\sem{t}$\\
$\isrel{\Sigma}{R}{a}{a'}$   & $\sem{R}\sq\sem{a}\times\sem{a'}$ \\
$\isprop{\Sigma}{\Gamma}{F}$ & $\sema{F}\in\{0,1\}$ \\
\hline
\end{tabular}
\caption{Semantics of Judgments}\label{fig:semantics}
\end{center}
\vspace{-3em}
\end{wrapfigure}

A $\Sigma$-model assigns to every symbol $s$ in $\Sigma$ a denotation. The formal definition is given in Def.~\ref{def:model}. Relative to a fixed model $M$ (which we suppress in the notation), each well-formed expression has a well-defined denotational semantics, given by the interpretation function $\sem{-}$. The semantics of propositions and queries in context $\Gamma$ is relative to an assignment $\alpha$, which assigns values to all variables in $\Gamma$. An overview is given in Fig.~\ref{fig:semantics}. The formal definition is given in Def.~\ref{def:semantics}.

\begin{definition}[Models]\label{def:model}
A $\Sigma$-model $M$ assigns to every $\Sigma$-symbol $s$ a denotation $s^M$ such that
\begin{itemize}
 \item $a^M$ is a set for $a:\btp$
 \item $c^M\sq\sem{a}$ for $c<a$
 \item $r^M\sq\sem{a}\times\sem{a'}$ for $r<\fp{a,a'}$
 \item $f^M:\sem{T_1}\times\ldots\times\sem{T_n}\arr\sem{T}$ for $f:\fp{T_1,\ldots,T_n}\arr T$
 \item $p^M:\sem{T_1}\times\ldots\times\sem{T_n}\arr\{0,1\}$ for $p:\fp{T_1,\ldots,T_n}\arr \prop$
\end{itemize}
\end{definition}

\begin{definition}[Semantics]\label{def:semantics}
Given a $\Sigma$-model $M$, the interpretation function $\sem{-}$ is defined as follows.

\noindent
Semantics of types:
\begin{itemize}
 \item $\sem{\qp{a_1,\ldots,a_n}}$ is the cartesian product $a_1^M\times \ldots\times a_n^M$
 \item $\sem{\set{t}}$ is the power set of $\sem{t}$
\end{itemize}

\noindent
Semantics of relations:
\begin{itemize}
  \item $\sem{r}=r^M$
  \item $\sem{R^{-1}}$ is the dual/inverse relation of $\sem{R}$, i.e., the set $\{(u,v)\,|\,(v,u)\in\sem{R}\}$
  \item $\tc{R}$ is the transitive closure of $\sem{R}$
  \item $R;R'$ is the composition of $\sem{R}$ and $\sem{R'}$,\\ i.e., the set $\{(u,w)|\mexists v\msuchthat (u,v)\in\sem{R},\;(v,w)\in\sem{R'}\}$
  \item $R\cup R'$, $R\cap R'$, and $R\sm R'$ are interpreted in the obvious way using the union, intersection, and difference of sets
\end{itemize}

\noindent
Semantics of propositions under an assignment $\alpha$:
\begin{itemize}
\item $\sema{p(Q_1,\ldots,Q_n)}=p^M(\sema{Q_1},\ldots,\sema{Q_n})$
\item $\sema{\neg F}=1 \tb\miff\tb \sema{F}=0$
\item $\sema{F\wedge F'}=1 \tb\miff\tb \sema{F}=1$ and $\sema{F'}=1$
\item $\sema{\forall x\in Q.F(x)}=1 \tb\miff\tb \sema[\alpha,\sub{x}{u}]{F(x)}=1 \tb \mforall u\in\sema{Q}$
\end{itemize}

\noindent
Semantics of queries $\ofSG{Q}{T}$ under an assignment $\alpha$:
\begin{itemize}
  \item $\sema{c}=c^M$
  \item $\sema{x}=\alpha(x)$
  \item $\sema{f(Q_1,\ldots,Q_n)}=f^M(\sema{Q_1},\ldots,\sema{Q_n})$
  \item $\sema{\rel{R}{Q}}=\{u\in\sem{a'}\,|\,(\sema{Q},u)\in\sem{R}\}$ for a relation $\isrel{\Sigma}{R}{a}{a'}$ and a query $\ofSG{Q}{a}$
   \\
  informally, $\sema{\rel{R}{Q}}$ is the image of $\sema{Q}$ under $\sem{R}$
  \item $\sema{\bigcup_{x\in Q}Q'(x)}$ is the union of all sets $\sema[\alpha,\sub{x}{u}]{Q'(x)}$ where $u$ runs over all elements of $\sema{Q}$
  \item $\sema{\{x\in Q|F(x)\}}$ is the subset of $\sema{Q}$ containing all elements $u$ for which $\sema[\alpha,\sub{x}{u}]{F(x)}=1$
\end{itemize}
\end{definition}

\begin{remark}\label{rem:finite}
It is easy to prove that if all concept and relation symbols are interpreted as finite sets and if all function symbols with result type $\set{t}$ always return finite sets, then all well-formed queries of type $\set{t}$ denote a \emph{finite} subset of $\sem{t}$.
Moreover, if the interpretations of the function and predicate symbols are computable functions, then the interpretation of queries is computable as well.
This holds even if base types are interpreted as infinite sets.
\end{remark}

\subsection{Predefined Symbols}

\begin{wrapfigure}{r}{7cm}
\vspace{-5em}
\begin{center}
\begin{tabular}{|l|l|l|}
\hline
Symbol & Type & Semantics \\
\hline
$\{\_\}$   & $:t\arr \set{t}$          & the singleton set \\
$\_\doteq\_$ & $:\fp{t, t}\arr\prop$       & equality \\
$\_\in\_$    & $:\fp{t,\set{t}}\arr\prop$ & elementhood \\
\hline
\end{tabular}
\caption{Predefined Symbols}\label{fig:predefined}
\end{center}
\vspace{-3em}
\end{wrapfigure}

We use a number of predefined function and predicate symbols as given in Fig.~\ref{fig:predefined}. These are assumed to be implicitly declared in every signature, and their semantics is fixed. All of these symbols are overloaded for all simple types $t$. Moreover, we use special notations for them.

All of this is completely analogous to the usual treatment of equality as a predefined predicate symbol in first-order logic. The only difference is that our slightly richer type system calls for a few additional predefined symbols.

It is easy to add further predefined symbols, in particular equality of sets (which, however, may be inefficient to decide) and binary union of queries.
We omit these here for simplicity.

\subsection{Definable Queries}

Using the predefined symbols, we can define a number of further useful query formation operators:

\begin{example}\label{ex:definable}
Using the singleton symbol $\{\_\}$, we can define for $\ofSG{Q}{\set{t}}$ and $\oftype{\Sigma}{\Gamma,x:t}{q(x)}{t'}$
 \[\{q(x)\,:\,x\in Q\} \;:=\;\bigcup_{x\in Q} \{q(x)\} \tb \mathrm{of\;type}\; \set{t'}.\]

It is easy to show that, semantically, this is the replacement operator, i.e., $\sema{\{q(x)\,:\,x\in Q\}}$ is the set containing exactly the elements $\sema[\alpha,\sub{x}{u}]{q(x)}$ for any $u\in\sema{Q}$.
\end{example}

\begin{example}[SQL-style Queries]\label{ex:sql}
For a query $\oftype{\Sigma}{}{Q}{\set{\qp{a_1,\ldots,a_N}}}$, natural numbers $n_1,\ldots,n_k\in\{1,\ldots,N\}$, and a proposition $\isprop{\Sigma}{x_1:a_1,\ldots,x_N:a_N}{F(x_1,\ldots,x_n)}$, we write
\[\mathbf{select}\;n_1,\ldots, n_k\; \mathbf{from}\;Q\;\mathbf{where}\;F(1,\ldots,N)\]
for the query
\[\{\qpi{x_{n_1},\ldots,x_{n_k}} \; : \; x\in \{y \in Q \,|\, F(y_1,\ldots,y_N)\}\} \]
of type $\set{\qp{a_{n_1},\ldots,a_{n_k}}}$.
\end{example}

\begin{example}[XQuery-style Queries]\label{ex:xquery}
For queries $\oftype{\Sigma}{}{Q}{\set{a}}$ and $\oftype{\Sigma}{x:a}{q'(x)}{a'}$ and $\oftype{\Sigma}{x:a,y:a'}{Q''(x,y)}{\set{a''}}$, and a proposition $\isprop{\Sigma}{x:a,y:a'}{F(x,y)}$, we write
\[\mathbf{for}\;x\;\mathbf{in}\;Q \; \mathbf{let}\;y = q'(x)\;\mathbf{where}\;F(x,y)\; \mathbf{return}\;Q''(x,y)\]
for the query
\[\bigcup_{z\in P} Q''(z_1,z_2) \tb\mwith\tb P:= \big\{z \in \{\qpi{x,q'(x)}\,:\,x\in Q\} \;|\; F(z_1,z_2)\big\}\]
of type $\set{a''}$.
\end{example}

\begin{example}[DL-style Queries]\label{ex:dl}
For a relation $\isrelS{R}{a}{a'}$, a concept $c<a$, and a query $\oftype{\Sigma}{}{Q}{\set{a'}}$, we write
  $\square^c R.Q$
for the query
 $\{x\in c \,|\,\forall y\in \rel{R}{x}.y\in Q\}$
of type $\set{a}$.

Note that, contrary to the universal restriction $\square R.Q$ in description logic, we have to restrict the query to all $x$ of concept $c$ instead of querying for all $x$ of type $a$. This makes sense in our setting because we assume that we can only iterate efficiently over concepts but not over (possibly infinite!) base types.

However, this is not a loss of generality: individual signatures may always couple a base type $a$ with a concept $\conc{is}_a$ such that $\sem{\conc{is}_a}=\sem{a}$.
\end{example}

\section{Querying MMT Libraries}\label{sec:mmtquery}
  \begin{figure}[ht]
\begin{center}
\begin{tabular}{|ll|p{6cm}|}
\hline
\multicolumn{2}{|l|}{Declaration} & Intuition \\
\hline
\multicolumn{3}{|c|}{Base types} \\
\hline
$\mmturi$   & $:\btp$               & URIs of {\mmt} declarations \\
$\mmtobj$   & $:\btp$               & {\mmt} (\openmath) objects \\
$\mmtxml$   & $:\btp$               & XML elements \\
\hline
\multicolumn{3}{|c|}{Concepts} \\
\hline
$\conc{theory}$   & $<\mmturi$ & theories \\
$\conc{view}$     & $<\mmturi$ & views \\
$\conc{constant}$ & $<\mmturi$ & constants \\
$\conc{style}$    & $<\mmturi$ & styles \\
\hline
\multicolumn{3}{|c|}{Relations} \\
\hline
$\conc{includes}$ & $<\fp{\mmturi,\mmturi}$ & inclusion between theories \\
$\conc{declares}$ & $<\fp{\mmturi,\mmturi}$ & declarations in a theory \\
$\conc{domain}$   & $<\fp{\mmturi,\mmturi}$ & domain of structure/view \\
$\conc{codomain}$ & $<\fp{\mmturi,\mmturi}$ & codomain of structure/view \\
\hline
\multicolumn{3}{|c|}{Functions} \\
\hline
$\typeOF$   & $:\mmturi\arr\opt{\mmtobj}$ & type of a constant \\
$\defOF$    & $:\mmturi\arr\opt{\mmtobj}$ & definiens of a constant \\
$\typeof$   & $:\fp{\mmturi,\mmtobj}\arr\opt{\mmtobj}$ & type inference relative to a theory \\
$\subobjat{p}$ & $:\mmtobj\arr\opt{\mmtobj}$ & argument at position $p$ \\
$\subobjhead$  & $:\fp{\mmtobj,\mmturi}\arr\set{\mmtobj}$ & subobjects with a certain head \\
$\unify$    & $:\mmtobj\arr\set{\qp{\mmturi,\mmtobj,\mmtobj}}$ & all objects that unify with a given one \\
$\render$   & $:\fp{\mmturi,\mmturi}\arr\opt{\mmtxml}$ & rendering of a declaration using a certain style \\
$\render$   & $:\fp{\mmtobj,\mmturi}\arr\opt{\mmtxml}$ & rendering of an object using a certain style \\
$u$         & $:\mmturi$               & literals for {\mmt} URIs $u$ \\
$o$         & $:\mmtobj$      & literals for {\mmt} objects $o$ \\
\hline
\multicolumn{3}{|c|}{Predicates} \\
\hline
$\occurs$& $:\fp{\mmturi,\mmtobj}\arr\prop$ & occurs in \\
\hline
\end{tabular}
\caption{The QMT Signature for {\mmt}}\label{fig:signature}
\end{center}
\vspace{-2em}
\end{figure}

We will now fix an {\mmt}-specific signature $\Sigma$ that customizes QMT with the {\mmt} ontology as well as with several functions and predicates based on the {\mmt} specification. The declarations of $\Sigma$ are listed in Fig.~\ref{fig:signature}.

For simplicity, we avoid presenting any details of {\mmt} and refer to \cite{RK:mmt:10} for a comprehensive description.
For our purposes, it is sufficient to know that {\mmt} organizes mathematical knowledge in a simple ontology that can be seen as the fragment of {\sc OMDoc} pertaining to formal theories. We will explain the necessary details below when explaining the respective $\Sigma$-symbols.

An {\mmt} \defemph{library} is any set of {\mmt} declarations (not necessarily well-typed or closed under dependency).
We will assume a fixed library $L$ in the following.
Based on $L$, we will define a model $M$ by giving the interpretation $s^M$ for every symbol $s$ listed in Fig.~\ref{fig:signature}.

\paragraph{Base Types}
We use three base types.
Firstly, every {\mmt} declaration has a globally unique \defemph{canonical identifier}, its {\mmt} URI. We use this to define $\mmturi^M$ as the set of all {\mmt} URIs declared in $L$.

$\mmtobj^M$ the set of all {\openmath} objects that can be formed from the symbols in $\mmturi^M$.
In order to handle objects with free variables conveniently, we use the following convention: All objects in $\mmtobj^M$ are technically closed; but we permit the use of a special binder symbol $\mathtt{free}$, which can be used to formally bind the free variables. This has the advantage that the context of an object, which may carry, e.g., type attributions, is made explicit.
Using general {\openmath} objects means that the type $\mmtobj$ is subject to exactly $\alpha$-equality and attribution flattening, the only equalities defined in the {\openmath} standard.
The much more difficult problem of queries relative to a stronger equality relation remains future work.

The remaining base type $\mmtxml$ is a generic container for any non-{\mmt} \defemph{XML data} such as HTML or presentation MathML. Thus, $\mmtxml^M$ is the set of all XML elements.
This is useful because the {\mmt} API contains several functions that return XML.

\paragraph{Ontology}
For simplicity, we restrict attention to the most important notions of the {\mmt} ontology; adding the remaining notions is straightforward. The ontology only covers the {\mmt} declarations, all of which have canonical identifiers. Thus, all concepts refine the type $\mmturi$, and all relations are between identifiers.

Among the {\mmt} \defemph{concepts}, \defemph{theories} are used to represent logics, theories of a logic, ontologies, type theories, etc.
They contain \defemph{constants}, which represent function symbols, type operators, inference rules, theorems, etc.
Constants may have {\openmath} \defemph{objects} \cite{openmath} as their type or definiens.
Theories are related via theory morphisms called \defemph{views}. These are truth-preserving translations from one theory to another and represent translations and models.
Theories and views together form a multi-graph of theories across which theorems can be shared.
Finally, \defemph{styles} contain notations that govern the translation from content to presentation markup.

{\mmt} theories, views, and styles can be structured by a strong module system. The most important modular construct is the $\conc{includes}$ relation for explicit imports. The $\conc{declares}$ relation relates every theory to the constants it declares; this includes the constants that are not explicitly declared in $L$ but induced by the module system.
Finally, two further relations connect each view to its $\conc{domain}$ and $\conc{codomain}$.

All concepts and relations are interpreted in the obvious way. For example, the set $\conc{theory}^M$ contains the {\mmt} URIs of all theories in $L$.

\paragraph{Function and Predicate Symbols}
Regarding the function and predicate symbols, we are very flexible because a wide range of operations can be defined for {\mmt} libraries.
In particular, every function implemented in the {\mmt} API can be easily exposed as a $\Sigma$-symbol.
Therefore, we only show a selection of symbols that showcase the potential.

In Sect.~\ref{sec:query}, we have deliberately omitted \defemph{partial function symbols} in order to simplify the presentation of our language. However, in practice, it is often necessary to add them. For example, $\defOF^M$ must be a partial function because (i) the argument might not be the {\mmt} URI of a constant declaration in $L$, or (ii) even if it is, that constant may be declared without a definiens.
The best solution for an elegant treatment of partial functions is to use option types $\mathit{opt}(t)$ akin to set types $\set{t}$.
However, for simplicity, we make $\sem{-}$ a partial function that is undefined whenever the interpretation of its argument runs into an undefined function application.
This corresponds to the common concept of queries returning an error value.

The partial \defemph{functions} $\typeOF^M$ and $\defOF^M$ take the identifier of a constant declaration and return its type or definiens, respectively.
They are undefined for other identifiers.

The partial function $\typeof^M(u,o)$ takes an object $o$ and returns its dynamically inferred type. It is undefined if $o$ is ill-typed.
Since {\mmt} does not commit to a type system, the argument $u$ must identify the type system (which is represented as an {\mmt} theory itself).
If $O$ is a binding object of the form $\mathtt{OMBIND}(\mathtt{OMS}(\mathtt{free}),\Gamma,o')$, the type of $o'$ is inferred in context $\Gamma$.

$\subobjat{p}$ is a family of function symbols indexed by a natural number $p$. $p$ indicates the position of a direct subobject (usually an argument), and $\subobjat{p}^M(o)$ is the subobject of $o$ at position $p$. In particular, $\subobjat{i}^M(\mathtt{OMA}(f,a_1,\ldots,a_n))=a_i$. Note that arbitrary subobjects can be retrieved by iterating $\subobjat{p}$.
Similarly, $\subobjhead^M(o,h)$ is the set of all subobjects of $o$ whose head is the symbol with identifier $h$. In particular, the head of $\mathtt{OMA}(\mathtt{OMS}(h),a_1,\ldots,a_n)$ is $h$. 
In both cases, we keep track of the free variables, e.g., $\subobjat{2}^M(\mathtt{OMBIND}(b,\Gamma,o))=\mathtt{OMBIND}(\mathtt{OMS}(\mathtt{free}),\Gamma,o)$ for $b\neq\mathtt{OMS}(\mathtt{free})$.

%

$\unify^M(O)$ performs an object query: It returns the set of all tuples $\qpi{u,o,s}$ where $u$ is the {\mmt} URI of a declaration in $L$ that contains an object $o$ that unifies with $O$ using the substitution $s$.
Here we use a purely syntactic definition for unifiability of {\openmath} objects.

$\render^M(o,u)$ and $\render^M(d,u)$ return the presentation markup dynamically computed by the {\mmt} rendering engine.
This is useful because the query and the rendering engine are often implemented on the same remote server. Therefore, it is reasonable to compute the rendering of the query results, if desired, as part of the query evaluation. Moreover, larger signatures might provide additional functions to further operate on the presentation markup.
$\render$ is overloaded because we can present both {\mmt} declarations and {\mmt} objects. In both cases, $u$ is the {\mmt} URI of the style providing the notations for the rendering.

The \defemph{predicate} symbol $\occurs$ takes an object $O$ and an identifier $u$, and returns true if $u$ occurs in $O$.

Finally, we permit \defemph{literals}, i.e., arbitrary URIs and arbitrary {\openmath} objects may be used as nullary constants, which are interpreted as themselves (or as undefined if they are not in the universe). This is somewhat inelegant but necessary in practice to form interesting queries.
A more sophisticated QMT signature could use one function symbol for every {\openmath} object constructor instead of using {\openmath} literals.

\begin{example}\label{ex:graph}
An {\mmt} theory graph is the multigraph formed by using the theories as nodes and all theory morphisms between them as edges. The components of the theory graph can be retrieved with a few simple queries.

Firstly, the set of theories is retrieved simply using the query $\conc{theory}$.
Secondly, the theory morphisms are obtained by two different queries:
\[\begin{array}{ll}
\mathrm{views} & \{\qpi{v,x,y} \,:\, v\in \conc{view},\;x\in \rel{\conc{domain}}{v},\;y\in \rel{\conc{domain}}{v}\} \\
\mathrm{inclusions} & \bigcup_{y\in \conc{theory}} \{\qpi{x,y} \,:\, x\in\rel{\tc{\conc{includes}}}{y}\}
\end{array}\]
The first one returns all view identifiers with their domain and codomain. Here we use an extension of the replacement operator $\{\_:\_\}$ from Ex.~\ref{ex:definable} to multiple variables. It is straightforward to define in terms of the unary one.
The second query returns all pairs of theories between which there is an inclusion morphism.
\end{example}

\begin{example}\label{ex:witness}
Consider a constant identifier $\exists I$ for the introduction rule of the existential quantifier from the natural deduction calculus. It produces a constructive existence proof of $\exists x.P(x)$; it takes two arguments: a witness $w$, and a proof of $P(w)$.
Moreover, consider a theorem with identifier $u$. Recall that using the Curry-Howard representation of proofs-as-objects, a theorem $u$ is a constant, whose type is the asserted formula and whose definiens is the proof.

Then the following query retrieves all existential witnesses that come up in the proof of $u$:
\[\{\subobjat{1}(x) \,:\, x\in \subobjhead(\defOF(u),\exists I)\}\]
Here we have used the replacement operator introduced in Ex.~\ref{ex:definable}.
\end{example}

\begin{example}[Continuing Ex.~\ref{ex:witness}]\label{ex:inference}
Note that when using $\exists I$, the proved formula $P$ is present only implicitly as the type of the second argument of $\exists I$.
If the type system is given by, for example, $\mathit{LF}$ and type inference for $\mathit{LF}$ is available, we can extend the query from Ex.~\ref{ex:witness} as follows:

\[\{\qpi{\subobjat{1}(x),{\typeof(\mathit{LF},\subobjat{2}(x))}} \,:\, x\in \subobjhead(\defOF(u),\exists I)\}\]

This will retrieve all pairs $(w,P)$ of witnesses and proved formulas that come up in the proof of $u$.
\end{example}

\section{Implementation}\label{sec:impl}
  We have implemented QMT as a part of the {\mmt} API.
The implementation includes a concrete XML syntax for queries and an integration with the {\mmt} web server, via which the query engine is exposed to users.
The server can run as a background service on a local machine as well as a dedicated remote server.
Sources, binaries, and documentation are available at the project web site \cite{project:mmt}.

The {\mmt} API already implements the {\mmt} ontology so that appropriate indices for the semantics of all concept and relation symbols are available.
Indices scale well because they are written to the hard drive and cached to memory on demand.
With two exceptions, the semantics of all function and predicate symbols is implemented by standard {\mmt} API functions.

The semantics of $\unify$ is computed differently: A substitution tree index of the queries library is maintained separately by an installation of MathWebSearch \cite{mathwebsearch}.
Thus, QMT automatically inherits some heuristics of MathWebSearch, such as unification up to symmetry of certain relation symbols.
MathWebSearch and query engine run on the same machine and communicate via HTTP.

Another subtlety is the semantics of $\typeof$. The {\mmt} API provides a plugin interface, through which individual type systems can be registered; the first argument to $\typeof^M$ is used to choose an applicable plugin.
In particular, we provide a plugin for the logical framework LF \cite{lf}, which handles type inference for any type system that is formalized in LF; this covers all type systems defined in the LATIN library \cite{CHKMR:latinabs:11} and thus also applies to our imports of the Mizar \cite{mizar} and TPTP libraries \cite{tptp}.

Query servers for individual libraries can be set up easily.
In fact, because the {\mmt} API abstracts from different backends, queries automatically return results from all libraries that are registered with a particular instance of the {\mmt} API.
This permits queries across libraries, which is particularly interesting if libraries share symbols. Shared symbols arise, for example, if both libraries use the standard OpenMath CDs where possible or if overlap between the libraries' underlying meta-languages is explicated in an integrating framework like the LATIN atlas \cite{CHKMR:latinabs:11}.


\begin{example}
The LATIN library \cite{CHKMR:latinabs:11} consists of over 1000 highly modularized LF signatures and views between them, formalizing a variety of logics, type theories, set theories, and related formal systems.
Validating the library and producing the index for the {\mmt} ontology takes a few minutes with typical desktop hardware; reading the index into memory takes a few seconds.
Typical queries as given in this paper are evaluated within seconds.

As an extreme example, consider the query $Q=\rel{\conc{Declares}}{\conc{theory}}$. It returns in less than a second the about 2000 identifiers that are declared in any theory.
The query $\bigcup_{x\in Q}\{\qpi{x,\typeOF(x)}\}$ returns the same number of results but pairs every declaration with its type. This requires the query engine to read the types of all declarations (as opposed to only their identifiers). If none of these are cached in memory yet, the evaluation takes about 4 minutes.
\end{example}


\section{Conclusion and Future Work}\label{sec:conc}
  We have introduced a simple, expressive query language for mathematical theories (QMT) that combines features of compositional, property, and object query languages.
QMT is implemented on top of the {\mmt} API; that provides any library that is serialized as {\mmt} content markup with a scalable, versatile querying engine out of the box.
As both {\mmt} and its implementation are designed to admit natural representations of any declarative language, QMT can be readily applied to many libraries including, e.g., those written in Twelf, Mizar, or TPTP.

Our presentation focused on querying \emph{formal} mathematical libraries.
This matches our primary motivation but is neither a theoretical nor a practical restriction.
For example, it is straightforward to add a base type for presentation MathML and some functions for it. MathWebSearch can be easily generalized to permit unification queries on presentation markup. This also permits queries that mix content and presentation markup, or content queries that find presentation results. Moreover, for presentation markup that is generated from content markup, it is easy to add a function that returns the corresponding content item so that queries can jump back and forth between them.

Similarly, we can give a QMT signature with base types for authors and documents (papers, book chapters, etc.) as well as relations like \texttt{author-of} and \texttt{cites}. It is easy to generate the necessary indices from existing databases and to reuse our implementation for them.
Moreover, with a relation \texttt{mentions} between papers and the type $\mmturi$ of mathematical concepts, we can combine content and narrative aspects in queries. An index for the \texttt{mentions} relation is of course harder to obtain, which underscores the desirability of mathematical documents that are annotated with content URIs. 

\bibliographystyle{alpha}
\newcommand{\etalchar}[1]{$^{#1}$}

\end{document}